\documentclass[journal]{IEEEtran}
\usepackage{cite}
\usepackage[normalem]{ulem}
\usepackage{ifpdf}
\usepackage{cite}
\usepackage{soul}
\usepackage{xcolor}
\usepackage{soul}
\usepackage{slashbox}
\usepackage{tikz,stanli,color}
\usepackage{amsmath,amssymb,amstext,stmaryrd,subfigure,array,multirow}
\usepackage{graphicx}
\usepackage[colorlinks=true, allcolors=blue]{hyperref}

\ifCLASSINFOpdf
\else
\fi
\usepackage{epsfig}
\usepackage{subfigure}
\usepackage{amsmath,amssymb,amstext} 
\usepackage{mathastext}

\begin{document}
\title{Mitosis Detection for Breast Cancer Pathology Images Using UV-Net}

\author{Seyed H. Mirjahanmardi$^1$, Samir Mitha$^1$, Salar Razavi$^1$, Susan Done$^{2,3}$, and April Khademi$^1$\\
$^1$Department of Electrical and Computer Engineering, Ryerson University, Toronto, ON, CAN\\
$^2$Department of Laboratory Medicine and Pathobiology, University of Toronto, Toronto, ON, CAN\\
$^3$Princess Margaret Cancer Centre, University Health Network, Toronto, ON, CAN}






\maketitle

\begin{abstract}
The difficulty of detecting mitosis and its similarity to non-mitosis objects has remained a challenge in computational pathology. The lack of publicly available data has added more complexity. Deep learning algorithms have shown potentials in mitosis detection tasks. However, they face challenges when applied to pathology images with dense medium and diverse dataset. This paper applies an optimized UV-Net architecture, developed to focus on mitosis details with high-resolution through feature preservation. Stain normalization methods are used to generalize the trained network. An F1 score of 0.6721 is achieved using this network.    
\end{abstract}

\section{Introduction}
Breast cancer is the second most commonly diagnosed cancer among women worldwide and histopathology has played a pivotal role in its diagnosis, prognostication, and treatment~\cite{shostak}. Traditionally, pathologists evaluate excised hematoxylin and eosin (H\&E) stained tissues under microscopes to analyze tissue microstructure, spatial nuclei configuration, and cellular morphology. Mitosis detection is one of the most critical parameters in cancer grading and prognosis \cite{saha}. Mitosis provides rich information about the tumor proliferation rate and its aggressiveness. Manual detection of mitosis by pathologists is time-consuming, laborious, and sometimes subjective to disagreement between experts. The advent of digital pathology wholeslide scanners has brought the potential to improve objectivity and turn-around-times (TATs). Deep learning methods have been showing promises in automating this process \cite{saha, li}. These algorithms range from recognition to segmentation based on convolutional neural networks. An example is Fast-RCNN based methods that are applied to provide mitosis segmentation \cite{ren, li}. 

\indent This paper uses our introduced architecture on nuclei detection, UV-Net, \cite{seyed} to focus on preserving high-resolution details in pathology images and identify mitosis. Architectures such as the widely-used U-Net \cite{ronneberger} that are composed of convolutional neural networks, may not be able to sufficiently recover details as successive convolutional layers and early maxpoolings remove high-resolution information and fine details that are important for quantifying mitosis (rare and small events). The UV-Net architecture preserves dense features through "V" blocks to retain the high-resolution details. Experiments are conducted on MIDOG challenge dataset. 
%
%
\section{Datasets and Materials}
The dataset used in this work are the wholeslide images (WSI) of breast tissue obtained from the MIDOG challenge 2021 \cite{Wilm}. The experimental dataset is comprised of 3840 RGB patches of size 512$\times$512, which contain either mitosis, hard-negative examples, or both. The annotated images were provided as the coordinates of boxes around each mitosis. This dataset is randomly split into 60$\%$, 20$\%$, and 20$\%$ for training, validation, and testing. For training purposes, the centroids of mitosis were computed and a Gaussian kernel was applied to assign the maximum probability to the center of mitosis while incorporating the texture of mitosis' surrounding environment \cite{seyed}. 
\subsection{Processing Pipeline and Frameworks}

\begin{figure}[t]
	\begin{center}
		\includegraphics[width=0.7\linewidth]{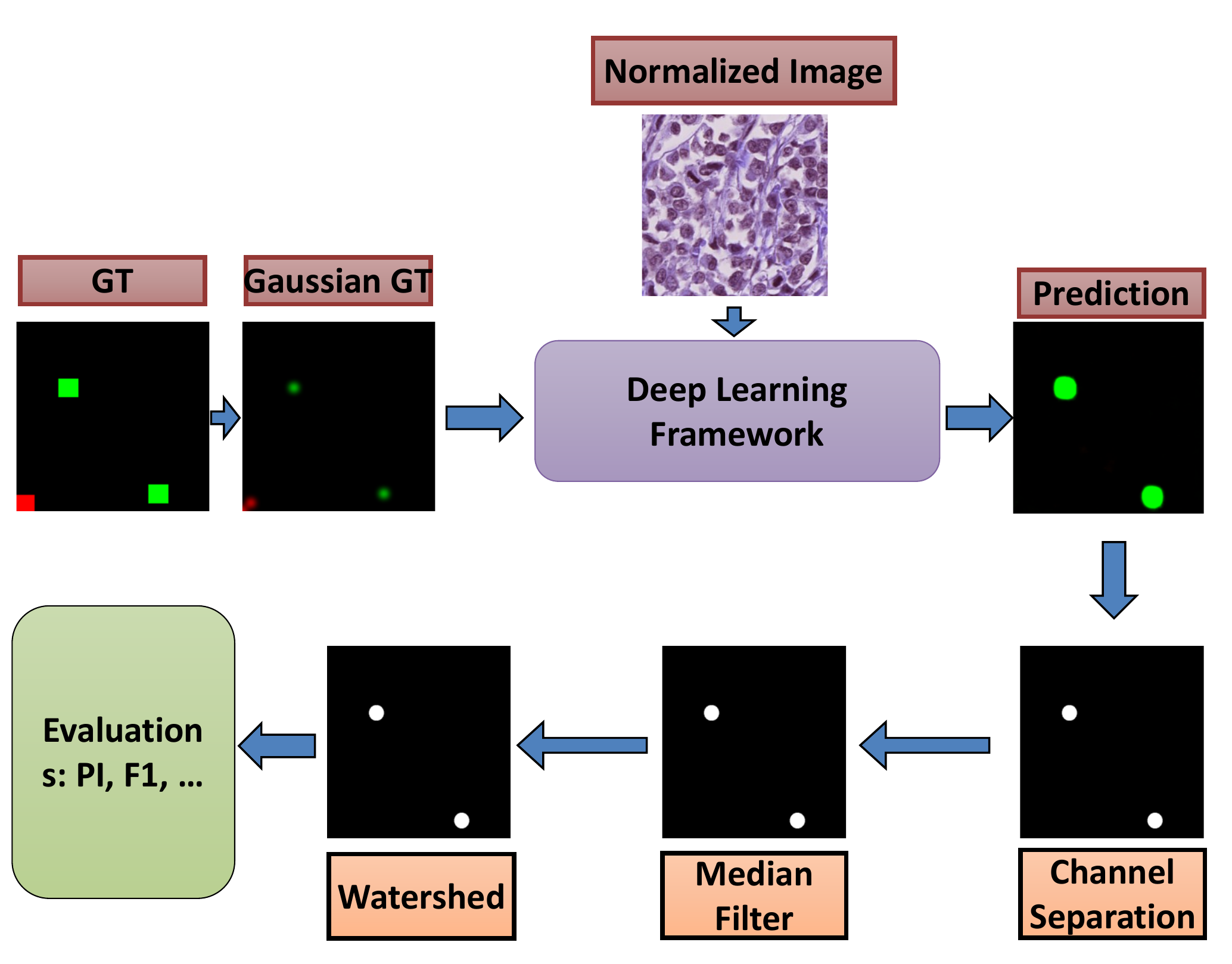}
		\caption{The process includes three steps: pre-processing (GT, Gaussian GT), deep learning framework, and post-processing (channel separation, Otsu thresholding, median filter, and watershed) }
		\label{fig-Pipeline}
	\end{center}
\end{figure}

The entire processing pipeline including the pre and post-processing is shown in Figure~\ref{fig-Pipeline}. The images are patched to size 512$\times$512, and the corresponding Gaussian GT images are created.  To enhance robustness and domain generalization, Macenko stain normalization was applied using the method provided in \cite{macenko}. A Huber loss function was used for all architectures to regress and predict the centroid of the mitosis. Data augmentation such as horizontal and vertical flips, as well as scaling are used. All experiments were conducted on the same machine with an NVIDIA GeForce RTX 2080 Ti. A total of 200 epochs were run with an Adam optimizer, batch size=4, and learning rate=10$^{-5}$.\\
\indent The predicted image is separated into two channels, one containing mitosis (green channel), and the other containing negative mitosis (red channel), the images are then post-processed. First, Otsu's thresholding is applied to each channel to convert the regressed prediction into a  binary representation. To remove small and irrelevant false positives, median filtering was applied. The watershed algorithm is then applied to disconnect the possible overlapped regions. The obtained results are then assessed to provided F1-score, precision and recall. While the algorithms is trained to classify both mitosis and negative mitosis, we focused on increasing the accuracy of mitosis detection. Thus, Figure~\ref{fig-Pipeline} only focuses on the green channel.

\section{Proposed Model: UV-Net}
We used our UV-Net architecture to detect mitosis. Figure~\ref{fig-Architecture} shows the full architecture, where 3$\times$3 convolutional layers used in U-Net are replaced by V-Blocks, inspired by the efficiency of dense connections. Each V-Block expands an input with n channels to output with 2n channels (creating a "V" shape) through four successive stages. Two hyperparameters, $f$ and $k$, are defined for each V-Block where they are equal to the number of input channels, and the output channels at the end of each stage, respectively. Figure~\ref{fig-Architecture}b shows a V-Block wherein $f=16$ and $k=4$. In each stage, the input feature is processed by a 1$\times$1 convolution with $f=16$ filters, then transformed to the output with $k=4$ filters. The output of this step is concatenated to the input, creating a matrix with 20 filters which are fed to the second stage. This process is repeated for a total of four times to generate an output with 2$\times$f filters. The successive concatenations prevent losing features obtained from earlier layers.

\newcommand{\subf}[2]{%
	{\small\begin{tabular}[t]{@{}c@{}}
			#1\\#2
	\end{tabular}}%
}
\begin{figure}
	\centering
	\begin{tabular}{c}

		\subf{\includegraphics[height=5cm]{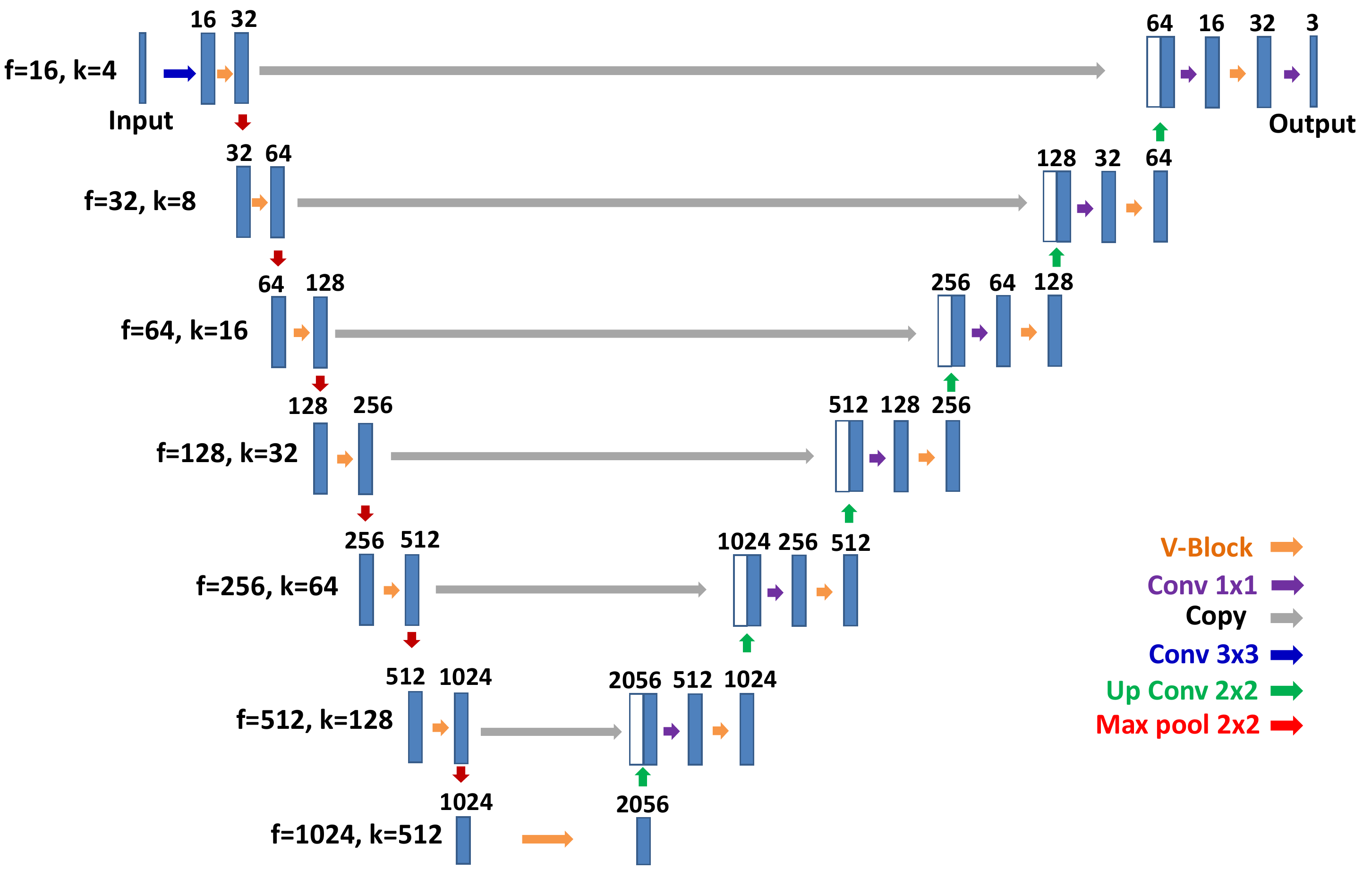}}
		{a}
		\\
		\subf{\includegraphics[height=2.6cm]{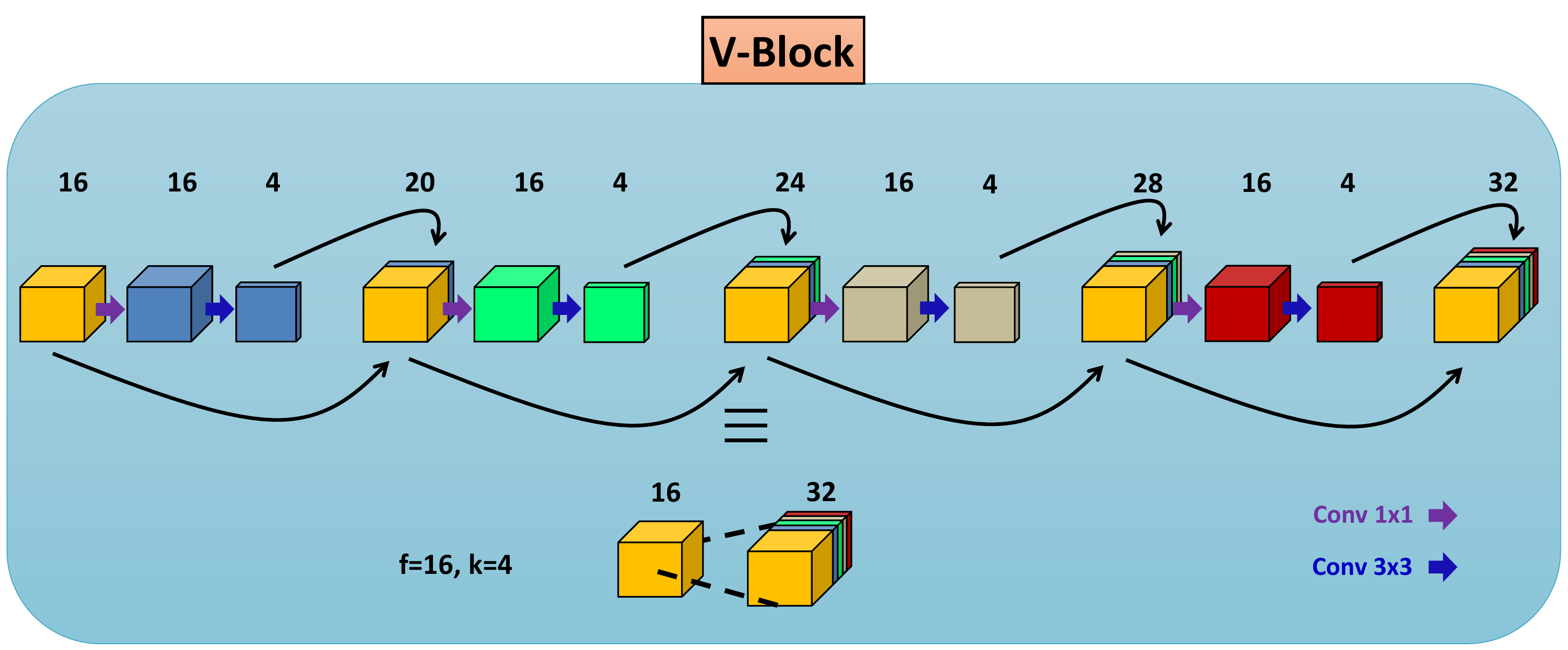}}
		{b}
		\\
	\end{tabular}
	\caption{UV-Net architecture with V-blocks including One V-block example. The output of each stage is concatenated with earlier outputs.}
	\label{fig-Architecture}
\end{figure}

\section{Results}
\begin{figure}[t]
	\begin{center}
		\includegraphics[width=0.6\linewidth]{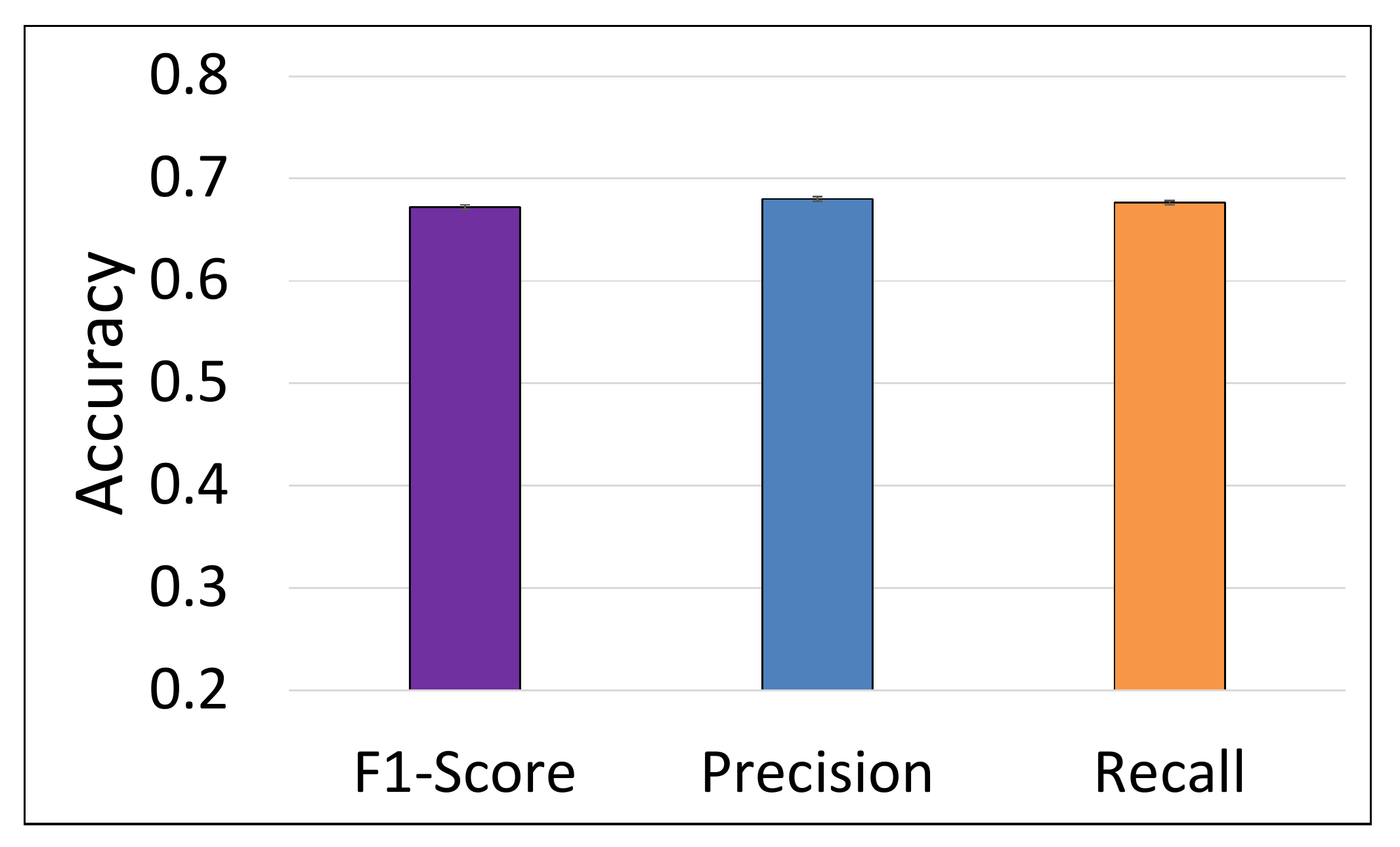}
		\caption{Accuracy results on unseen dataset including F1-Score, precision, and recall. }
		\label{fig-Results}
	\end{center}
\end{figure}

This section presents the results of UV-Net, tested on 768 unseen images of 512$\times$512 that have mitosis or hard-negative labels. Steps explained in Figure~\ref{fig-Pipeline} are followed to post-process the image and perform quantitative assessment. Figure~\ref{fig-Results} shows the result of UV-Net prediction including F1-score, precision, and recall. The obtained accuracy for F1-score, precision, and recall are 0.6721, 0.6800, and 0.6766, respectively.

%
\section{Conclusion}

This paper introduced an architecture referred to as UV-Net to focus on dense features and restore high-resolution details for mitosis detection across images from different scanners. UV-Net showed an F1-score of 0.6721 for the mitosis class using 768 test images.

\bibliographystyle{IEEEtran}
\bibliography{UV-Net}

\end{document}